\documentclass{optica-article}

\journal{oe}

\articletype{Research Article}

\usepackage{multirow}
\usepackage{lineno}
\usepackage{amsmath}

\begin{document}

\title{Characterizing the Dark Count Rate of a Large-Format MKID Array}

\author{Noah Swimmer\authormark{1,*}, W. Hawkins Clay\authormark{1}, Nicholas Zobrist\authormark{1}, Benjamin A. Mazin\authormark{1,$\dagger$}}

\address{\authormark{1}Department of Physics, University of California, Santa Barbara, California, 93107, USA}

\email{\authormark{*} nswimmer@ucsb.edu}
\homepage{\authormark{$\dagger$} http://mazinlab.org}



\begin{abstract}
We present an empirical measurement of the dark count rate seen in a large-format MKID array identical to those currently in use at observatories such as Subaru on Maunakea. This work provides compelling evidence for their utility in future experiments that require low-count rate, quiet environments such as dark matter direct detection. Across the bandpass from 0.946-1.534 eV (1310-808 nm) an average count rate of $(1.847\pm0.003)\times10^{-3}$ photons/pixel/s is measured. Breaking this bandpass into 5 equal-energy bins based on the resolving power of the detectors we find the average dark count rate seen in an MKID is $(6.26\pm0.04)\times10^{-4}$ photons/pixel/s from 0.946-1.063 eV and $(2.73\pm0.02)\times10^{-4}$ photons/pixel/s at 1.416-1.534eV. Using lower-noise readout electronics to read out a single MKID pixel we demonstrate that the events measured while the detector is not illuminated largely appear to be a combination of real photons, possible fluorescence caused by cosmic rays, and phonon events in the array substrate.  We also find that using lower-noise readout electronics on a single MKID pixel we measure a dark count rate of $(9.3\pm0.9)\times10^{-4}$ photons/pixel/s over the same bandpass (0.946-1.534 eV) With the single-pixel readout we also characterize the events when the detectors are not illuminated and show that these responses in the MKID are distinct from photons from known light sources such as a laser, likely coming from cosmic ray excitations.
\end{abstract}


\section{Introduction} \label{sec:intro}

Microwave Kinetic Inductance Detectors (MKIDs) are superconducting microresonators with the ability to measure individual photons\cite{day2003}. These light sensitive detectors can measure photon arrival times with microsecond precision and their unique detection mechanism also allows them to measure the energy of each incident photon. Since they utilize a different readout scheme than conventional image sensors such as charge-coupled devices (CCDs), MKIDs have the additional benefit of not being susceptible to read noise (stemming from the conversion of electrons from a CCD potential well into a voltage signal before being quantized and processed) or dark current (when thermal electrons accumulate in the potential well of the CCD and are counted as part of the signal upon readout despite their different origin)\cite{Szypryt2017, Zobrist2019, Fruitwala2020, Zobrist2021, Zobrist2022}. This, along with their photon counting ability makes MKIDs exceptional detectors for astronomy in photon starved regimes \cite{Steiger2021, Swimmer_2022, Steiger_2022}.

Current-generation MKIDs have been designed to be sensitive to photons in different ranges such as for ultraviolet, optical, and near-infrared (UVOIR) astronomy and X-ray detection. They also offer straightforward ways to tune their sensitivity to higher- or lower-energy bandpasses. This ability opens opportunities for MKIDs to be used as detectors for new physics applications such as the search for dark matter. In `photon-starved' regimes, where sources emit very few photons, it is of high importance to adequately characterize the performance of the detectors to ensure that each photon detected is `real', i.e. light from the astronomical source being observed rather than from errant sources such as thermal blackbody radiation from inside a cryostat.

In this work we aim to characterize ``dark counts'' measured by a large-format MKID array. These are events that are registered as photons by the detector when it is not exposed to a light source. These dark counts differ from those of conventional semiconductor detectors because of their origin. Semiconductor detectors register false counts due to dark current and read noise. Dark current is the generation of thermal electrons in the material that are captured by the detector's potential well and counted as part of the signal while read noise is the noise that is added to the measured signal from charge photon-to-voltage conversion and signal processing such as analog to digital conversion. In MKIDs false triggers may stem from noise in the room-temperature readout\cite{Fruitwala2020}, blackbody photons from the environment, or more complex sources. 

\section{MKID photon measurement}
\begin{figure}
    \centering
    \includegraphics[width=0.8\columnwidth]{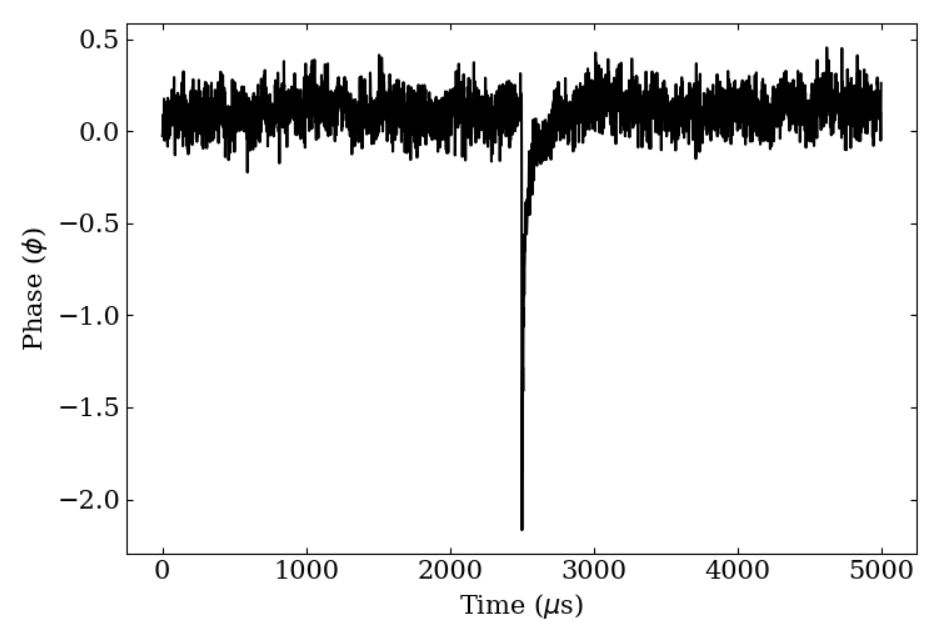}
    \caption{A single photon event measured by an MKID pixel with readout sampling at 0.8 MHz. The x-axis shows time in microseconds and y-axis shows the detector phase response measured in radians.}
    \label{fig:photon_timestream}
\end{figure}


\begin{figure}
    \centering
    \includegraphics[width=0.47\columnwidth]{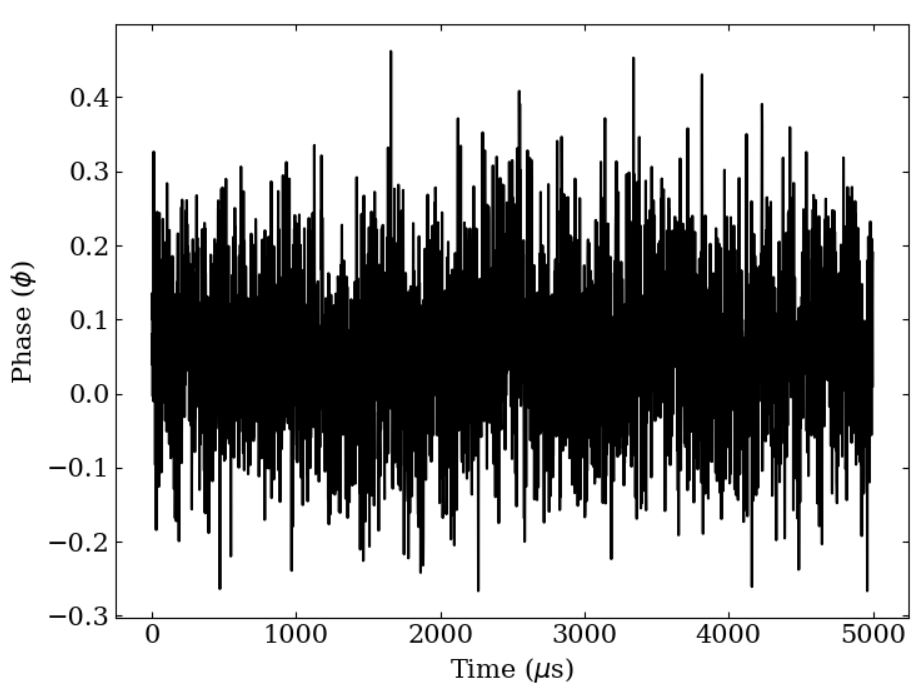}
    \includegraphics[width=0.468\columnwidth]{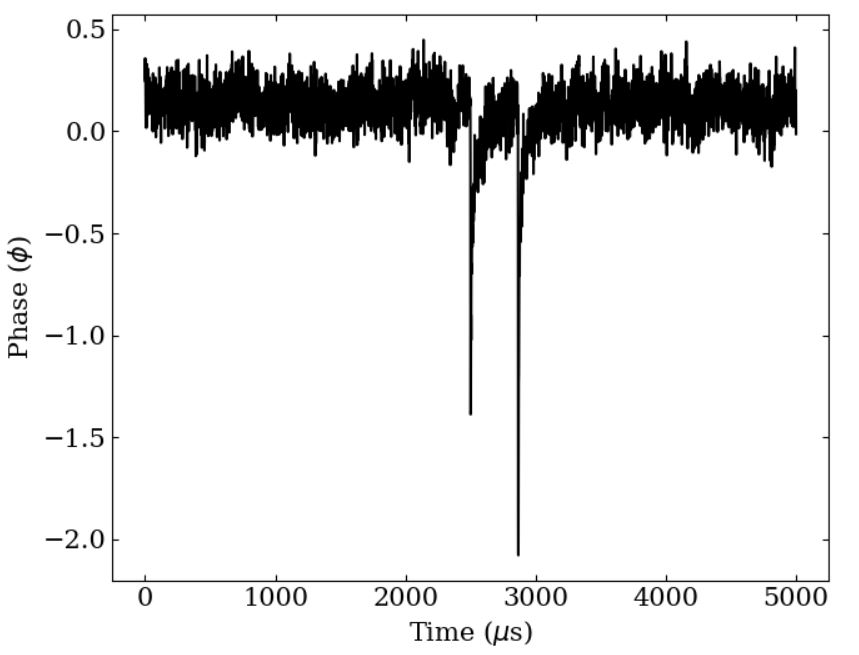}
    \caption{2 classes of `bad' photon events that can be triggered in the MKID readout. (Left) A phase time stream (phase measured in radians) taken when noise in the data caused the readout to trigger as if a photon hit the detector although clearly none did. (Right) A second photon arriving on the tail of a previous one. This contaminates the analysis of the first and, depending on the dead time, the readout may not trigger at all on the second.}
    \label{fig:bad_photon_triggers}
\end{figure}

Each MKID pixel -- ``pixel'' and ``detector'' may be used interchangeably -- is a superconducting LC-circuit which is excited at its resonant frequency by a probe tone using room temperature readout electronics. When an incident photon is absorbed by an MKID pixel, the energy from the photon breaks Cooper pairs in the resonator which causes the inductance of the resonator to increase. This increase in inductance is measured as a change in the phase of the microwave probe tone. The probe tone for each MKID pixel is typically sampled at 1 MHz, giving microsecond timing resolution \cite{Mazin2012, Szypryt2017, Fruitwala2020}.

For an MKID pixel to detect a photon, the photon event must cause the phase of the detector to increase beyond a minimum threshold. For a given pixel this threshold is calculated by measuring the phase (in radians) while not illuminated. This allows a phase noise to be measured, which we find does not typically exceed $\sigma_{\phi}\sim$0.15 radians in good-performing pixels. The threshold for the given resonator is then set to be 6$\sigma_{\phi}$ away from the average phase value -- typically 0 radians. In other terms, the threshold for each pixel is set to be
\begin{equation}
\label{eqn:threshold}
    \phi_{\rm Threshold}=\hat{\phi}+6\sigma_{\phi}\approx6\sigma_{\phi}
\end{equation}
where the right hand approximation holds when $\hat{\phi}\approx$0. This is the method by which each pixel's photon detection threshold is calculated for both electronic readout systems described in this paper (Sections \ref{sec:digitalreadout} and \ref{sec:analogreadout}). Since the phase noise is Gaussian and a 6$\sigma_{\phi}$ threshold is used, that means that there is less than a 1 in 500,000,000 chance that the phase will fluctuate above the threshold at any point when the phase is measured. The process for determining the threshold for a pixel is described in detail in \cite{SteigerAndBailey2022}. Additionally, each pixel's phase response to photons of different energies within the bandpass of interest is measured when the array is calibrated (Section \ref{sub:array_cal}). For the MKID resonators used in this experiment we find that the typical phase response for the lowest-energy photons in the bandpass (0.946 eV, $\lambda$=1310 nm) is $\phi_{\rm 0.946 eV}=-1.4\pm0.2$ radians and the typical phase response for the highest-energy photons (1.534 eV, $\lambda$=808 nm) is $\phi_{\rm 1.534 eV}=-2.2\pm0.3$ radians. This demonstrates that the phase noise is unlikely to ever swing sufficiently high to cause a resonator to trigger on an event within the calibrated bandpass.

It is also instructive to calculate the signal-to-noise ratio (SNR) for photon pulses seen by each resonator. The resolving power $\mathcal{R}=E/\Delta E\sim E/2.355\sigma_{\phi}$, where $E$ is the energy of a photon event and is proportional to the phase pulse height, meaning that the SNR $\sim E/\sigma_{\phi}$ (see Section \ref{sub:array_cal}). Rearranging, one finds that the SNR $\sim2.355\mathcal{R}$. This means that for a resonator with an $\mathcal{R}=4.5$ -- a typical value in this experiment -- the SNR for a photon pulse will be about 10.6, again demonstrating that a resonator's phase response to a photon will be significantly higher than the phase noise itself.

An example of a single photon event is shown in Figure \ref{fig:photon_timestream}. When a photon is measured the readout records the time at which the photon struck and the height of the peak of the pulse. The height of the pulse (which is proportional to the change in the resonator's phase) is related to the energy of the incident photon. Therefore, a photon measured using the MKID readout intrinsically gathers information about both the time a photon arrived and its energy. By performing an energy calibration with lasers of known wavelength, these measured pulse heights can be converted to energies of the incident photons within the calibrated range.

Figure \ref{fig:bad_photon_triggers} shows several types of events measured by a single detector that can contaminate a dataset. The first occurs when noise in the system causes the readout to register a photon when there is clearly none present. This may occur when there is a small random spike in the phase data or a very low energy photon hits the detector. These ``noise'' triggered events are removed in post-processing as their peak pulse heights fall outside of the calibrated energy bandpass and can therefore be removed when cutting photons that are outside this energy range. This cut will exclude any of the ``noise'' triggers that initially registered as a photon. The second is when multiple photons are caught riding on the tail of the initial photon registered by the readout. This can be mitigated by decreasing the detector `dead time', the time after a photon is measured before the readout can read a second photon. In the digital readout (Section \ref{sec:digitalreadout}) the first photon is counted while the second (and beyond) photon is ignored completely, while in the analog readout (Section \ref{sec:analogreadout}) one can manually calibrate these and split up timestreams that have multiple photon events within them, including both if desired or treating them similarly to the digital readout and not registering the rest.

\section{Experiment overview}

\subsection{Optics}
\begin{figure}
    \centering
    \includegraphics[width=\columnwidth]{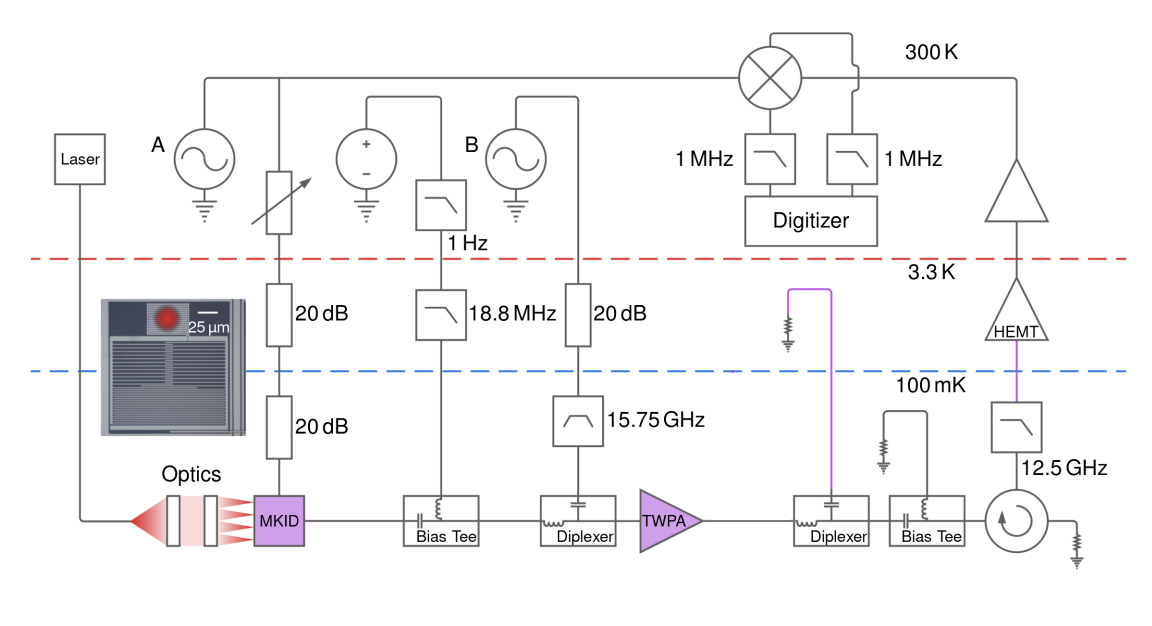}
    \caption{Schematic of the optical path from the laser at room temperature to the MKID array and of the idealized analog readout with parametric amplification. Laser light is carried down a multimode fiber that is inserted into the input port of a collimator. The collimated light is incident on a microlens array which then focuses the light onto each MKID pixel. The signal comes from synthesizer A, and the pump tone for the TWPA comes from synthesizer B. Each of the resistors to ground represents a 50$\Omega$ termination. (Inset) A single MKID pixel showing the ideal focus spot of light onto the photosensitive inductor portion of the detector.}
    \label{fig:dilution_refrigerator}
\end{figure}

The optical path to put light onto the MKID array is relatively simple. One of five lasers within the sensitivity range of the devices (808 nm, 920 nm, 980 nm, 1120 nm, 1310 nm) can be inserted into an integrating sphere which has an output to a multimode fiber at room temperature. The fiber is then inserted into a port in the dilution refrigerator where it is routed to directly to a collimator at the same temperature as the MKID device. The fiber has transmission greater than 90$\%$ at all of the laser wavelengths above. within the The collimator is oriented so the collimated light shines directly onto a microlens array (MLA) which focuses the light onto the photosensitive inductor of each MKID pixel. The collimator creates a spot greater in size than the MKID array, meaning that the distribution of light from the lasers is spatially uniform across the MKID pixels. A schematic of the optical path within the fridge is shown in Figure \ref{fig:dilution_refrigerator}, while the inset shows a spot of light focused on a single MKID pixel. 

To prevent potential stray photons from the inside of the dilution refrigerator from hitting the MKID array, a special lid was created so that the fiber collimator mounts directly to the box which houses the MKID array, preventing any unwanted photons from entering the window and hitting a detector.

\subsection{MKID array}
\label{sub:mkid_array}

\begin{figure}
    \centering
    \includegraphics[width=0.7\columnwidth]{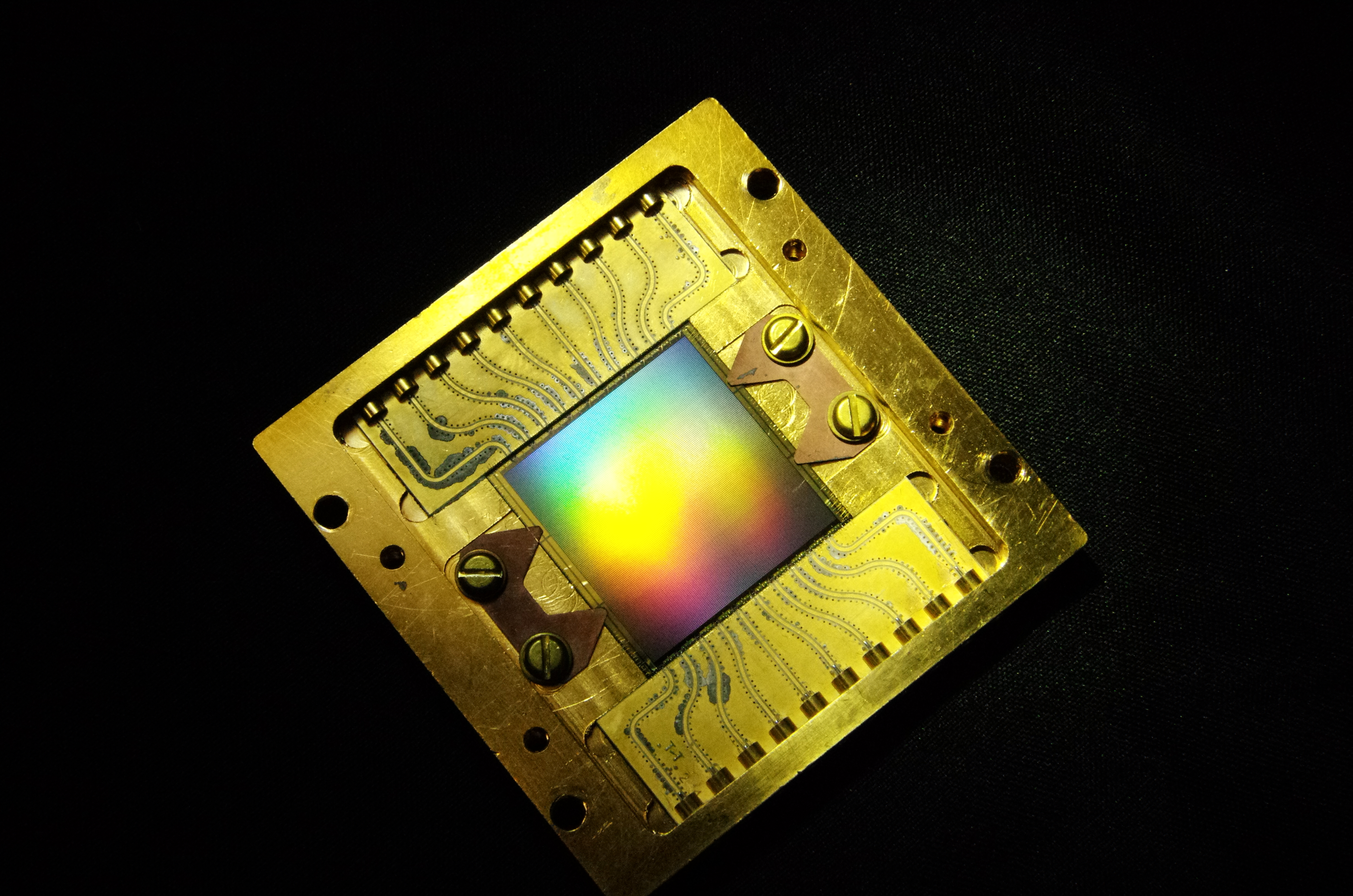}
    \caption{A 10-feedline MEC-style MKID array with 20,440 pixels. The lid with microlens array has been removed so that the MKID array itself can be seen. The MKID digital readout (Section \ref{sec:digitalreadout}) combined with the dilution refrigerator setup (Figure \ref{fig:dilution_refrigerator}) allow for up to 1 of the microwave feedlines (2044 MKID pixels) to be read out at a time. In an MKID instrument all 10 can be read out simultaneously.}
    \label{fig:mec_array}
\end{figure}

The MKID array used in this experiment has 10 microwave feedlines, each with 2044 MKID resonators coupled to it for a total of 20,440 pixels. This device has the same design as the MKID arrays currently in use with the MKID Exoplanet Camera (MEC, \cite{Walter2020}) which is currently commissioned at the Subaru Telescope on Maunakea. An image of a MEC-style array can be seen in Figure \ref{fig:mec_array}.

In the MEC instrument each of the 20,440 pixels can be read out simultaneously using the 2nd Generation MKID Digital Readout (see Section \ref{sec:digitalreadout} for further details)\cite{Fruitwala2020} which provides exquisite spatial resolution while time-tagging incident photons with microsecond accuracy and is theoretically capable of resolving each photon's energy to within several percent\cite{Mazin2012} although the resolving power may be degraded by noise from hard-to-control experimental sources such including electrical noise in the room-temperature readout electronics, magnetic fields and Johnson noise in the dilution refrigerator, two-level system noise in the detectors, and phase measurement noise. The resolving power may also be affected by strong radio frequency interference (RFI) from sources such as 5 GHz Wi-Fi that are close in frequency space to the resonant frequencies of the MKIDs (4-8 GHz). 

While mounted in an instrument, and MKID array has a lid to protect stray light that may be incident on it and a microlens array (MLA) which is designed to focus light onto the photosensitive region of each MKID resonator. Figure \ref{fig:mec_array} does not show the lid+MLA so that the array itself may be seen for clarity.

\subsection{Array calibration}
\label{sub:array_cal}

\begin{table}
    \centering
    \begin{tabular}{|| c |  c | c | c ||}
        \hline
        Energy (eV) & Wavelength (nm) & Median $\mathcal{R}_{1}$ & Median $\mathcal{R}_{2}$ \\
        \hline\hline
        0.946 & 1310 & 4.5 & 4.9 \\    
        1.107 & 1120 & 4.4 & 4.5 \\
        1.265 & 980 & 4.4 & 4.4 \\
        1.348 & 920 & 4.4 & 4.5 \\
        1.534 & 808 & 4.6 & 4.5 \\
        \hline
    \end{tabular}
    \caption{The median resolving power of all resonators measured at each calibration laser energy (wavelength). Columns 1 and 2 show the energy and wavelength of each laser used for calibration. Column 3 shows the median resolution values of all resonators taken prior to data collection while column 4 shows the same values taken 3 days after, following the conclusion of data collection.}
    \label{table:resolutions}
\end{table}

The resolving power of an MKID pixel for a given photon energy is given by $\mathcal{R}= E/\Delta E$. Here, $E$ is the energy of the incident photons -- typically from a monochromatic laser -- and $\Delta E$ is measured by determining the full width at half maximum (FWHM) of the distribution of photon energies recorded by the pixel. To measure the resolving power of each resonator, five lasers are shined across the array, one at a time. The responses of each resonator are recorded and used to find its resolving power at each of the different laser wavelengths\cite{SteigerAndBailey2022}.

819 MKID pixels were initially identified before any data collection. An energy calibration dataset was taken prior to and after data acquisition to assess the stability of each pixel's response to photons of a given energy. 

We require that each resonator was successfully energy calibrated in both datasets. This means that each pixel was marked by the energy calibration software as being successfully calibrated at each of the 5 laser energies. This cut removed 40 pixels, leaving 779 (95.1$\%$) of the original 819. This cut was made for two reasons. The first being that if the MKID Data Reduction Pipeline \cite{SteigerAndBailey2022} is not able to identify a resonator in one of the two calibration datasets, we cannot confirm that it stayed stable through the duration of the data collection. This was responsible for removing 14 of the 40 pixels cut. The second is that even if a resonator was identified in both calibration datasets, it must be able to be successfully energy calibrated at all 5 laser energies. We have seen that resonators that only pass 3 or 4 of the laser energies are worse performing and typically see many low-energy `noise triggered events' that make such pixels unreliable. This led to the other 26 of the 40 removed pixels being cut. The median $\mathcal{R}$ values at each energy are shown for each calibration dataset in Table \ref{table:resolutions}.

These 779 best performing pixels that remained after these cuts are those whose data will be used for the dark count analysis in Section \ref{sub:digitalreadout_collectionandreduction}. In Section \ref{sec:analogreadout} a single one of these resonators will be used to characterize the nature of the photons that are seen.

\subsection{Electronic readouts}
This investigation used 2 separate readout systems. The first readout system is the MKID Digital Readout described in \cite{Fruitwala2020} which used to read a substantial number of MKID pixels on one (or more) microwave feedline(s). The second system uses an analog readout scheme similar to that described in \cite{Zobrist2019,Zobrist2022} to read out a single MKID pixel. This system ensures lower readout noise and the ability to record photon phase timestreams . We will discuss the former in Section \ref{sec:digitalreadout} and the latter in Section \ref{sec:analogreadout}.

\section{MKID digital readout: many-detector measurement}
\label{sec:digitalreadout}

\subsection{Experimental setup} 
\label{sub:digitalreadout}
The full-array readout in this experiment used the second generation MKID digital readout\cite{Fruitwala2020} which is divided between 3 temperature stages where most of the large electronics are at room temperature. The remainder were housed in a BlueFors Dilution Refrigerator which cools them to 4K or 100 mK, depending on the component. The internal schematic for the fridge is seen in Figure \ref{fig:dilution_refrigerator}.

The room-temperature components are discussed in detail in \cite{Fruitwala2020} but are briefly described here for context and clarity. A 2nd-generation MKID readout board has a Xilinx Virtex-7 FPGA (Field Programmable Gate Array) controls Analog-to-Digital Converter/Digital-to-Analog Converter (ADC/DAC) boards which will generate (DAC) and take in (ADC) the in-phase (\textit{I}) and quadrature (\textit{Q}) components of each probe tone. It also contains ROACH-2 (Reconfigurable Open Architecture Computing Hardware) boards developed by CASPER (Collaboration for Astronomy Signal Processing and Electronics Research), which includes all of the core digital signal processing for the readout system\cite{Wethimer2011, Hickish2016}. This includes the channelization of probe tones, filtering, and photon pulse detection. An RF/IF board is also used for conversion of probe tones from the IF band (-1 to 1 GHz) up to the RF band (4-6 or 6-8 GHz) so that up to 1022 MKID resonators can be read out at their resonant frequencies. Each readout cartridge contains two sets of readout boards - one to read out pixels with resonant frequencies from 4-6 GHz and the other to read out pixels with resonant frequencies from 6-8 GHz - to provide the capability to read out all of the 2044 MKID pixels on a single microwave feedline. For this experiment we chose to only read out the higher-frequency half of an MKID feedline from 6-8 GHz to reduce potential noise from boards running on the same cartridge and to best match the bandpass of the parametric amplifier.

Two 4-meter RF coaxial cables are then attached between the readout cartridge and the dilution fridge to send the signal to the MKID array and carry the output back to the readout cartridge to be read out.

Internally, the signal is sent from room temperature to 100 mK using cryogenic RF coax cables that are heat sunk at intermediate temperature stages to reduce heat flow to the MKIDs. At the 4K stage there is a 20 dB attenuator to attenuate Johnson noise from the room temperature input. A second 20 dB attenuator is added at the 100 mK stage to further reduce Johnson noise from the 4K stage and to account for the amplification of the signals on the output side of the array. The probe tones are then sent through the MKID device, exciting the resonators of a single MKID array microwave feedline. On the output side the signal is sent through a traveling wave parametric amplifier (TWPA\footnote{Further work is being conducted to characterize the performance of a TWPA while reading out many MKID pixels and will be the subject of a future publication.}). This is a wideband (its maximal gain occurs between 6-8 GHz), high-power amplifier capable of reading out photon events with near quantum-limited amplifier noise \cite{Eom2012, Zobrist2019}. The signal is further amplified by a Low Noise Factory (LNF) High Electron Mobility Transistor (HEMT) amplifier at 4K. HEMT amplifiers are often used in cryogenic systems that require relatively high saturation power, significant dynamic range requirements, and a wide bandpass (for MKIDs, 4-8 GHz). The microwave signal is then sent back up to room temperature and routed back to the readout cartridge so the MKIDs may be measured.

Using the readout to monitor half of one MKID feedline enables up to 1022 pixels to be read out during this experiment. The number of functioning pixels compared to the total number of possible pixels is called the pixel yield. The current-generation 20,440 pixel arrays such as the one used in this experiment have yields of about 80$\%$. Each pixel in an MKID array is unique and can be rendered non-functioning for different reasons such as a piece of dust or residue landing on a pixel and shorting it to ground. Variation in the thickness of the resonators or their chemical composition can also cause them to move to unpredictable frequencies, colliding with another resonator in frequency space and making one or both unusable\cite{Walter2020}. The pixel yield of about 80$\%$ leads to 819 of the possible 1022 resonators being read out. These pixels will form the basis of our analysis of the dark count rate measured by the MKIDs. We note that this is the first simultaneous readout of a large MKID array with a parametric amplifier that we are aware of.

\subsection{Data collection and reduction}
\label{sub:digitalreadout_collectionandreduction}

Using the MKID second generation digital readout\cite{Fruitwala2020} each resonator from half of an MKID array feedline collected data with no light incident on the detectors for 86,750 seconds (24.1 hours in total) from 10-13 December 2020 while the dilution refrigerator was regulating the device temperature at 100 mK.

Notably, the data taken using the digital readout is useful because it allows hundreds to thousands of resonators to be read out simultaneously. This enables useful characterization of the data that can only be inferred via bulk properties of the array such as identification and rejection of cosmic ray hits (which contaminate the data) or `flashes' across the device (whose origins are currently unexplored but whose effects are nevertheless contaminating and must be removed).

\begin{figure}
    \centering
    \includegraphics[width=\columnwidth]{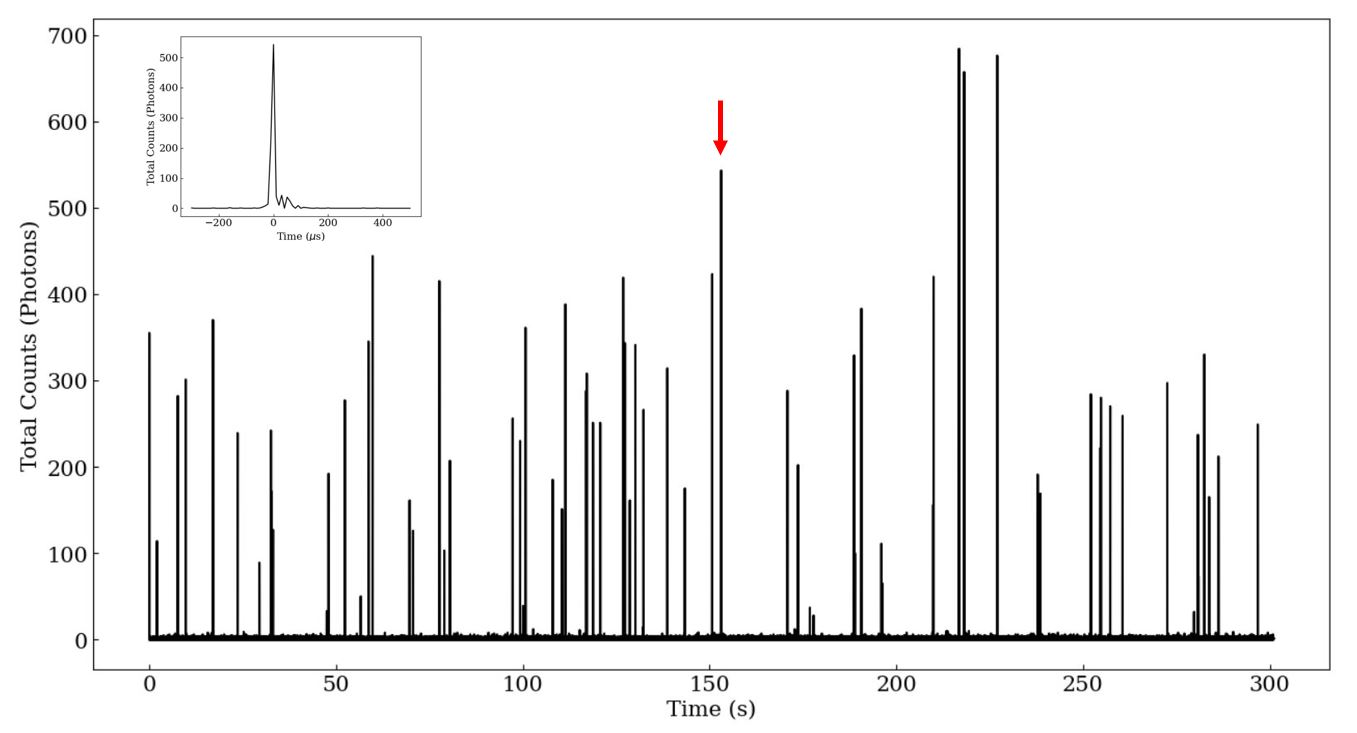}
    \caption{A timestream showing the number of counts across the 819 resonators read out using the MKID Digital Readout over a 300-second span. The timestream is created by binning the photon arrival times using 10 $\mu$s bins (i.e. all photons measured by the 819 resonators are first sorted by time then grouped into bins of 10 $\mu$s, meaning the y-axis shows the number of photons that hit the measured resonators within that short time range). The quiescent count rate is approximately 0 counts/second, but cosmic rays hitting the array and other potential system noise cause `flashes' where many pixels register a photon. The inset shows a single cosmic ray event (marked by the red arrow) for 200 $\mu$s before and 600 $\mu$s after the peak.}
    \label{fig:cosmics}
\end{figure}

The 86,750 seconds of data taken using the MKID Digital Readout were processed and reduced using the MKID Science Data Pipeline\cite{SteigerAndBailey2022}. The data were split into smaller chunks of time in order to create photon lists (the `base' format for MKID instrument data, see \cite{Steiger2021} for further details) of manageable size.

During this reduction, a cosmic ray calibration is performed by making a time-based coincidence veto based on significantly more photons being detected across the array in a short burst above the quiescent count rate. This mitigates cosmic ray events, which occurs when a high energy particle is absorbed by the MKID array. When an energetic particle absorbed in the substrate it may down-convert into a cloud of phonons that spreads out across the array, depositing energy into resonators as the phonons move away from the point of absorption. As energy is deposited into each resonator, that resonator will register a single spurious event. Since the energy from the energetic particle spreads so quickly over many resonators, this will cause a spike in the number of counts across the array for a short time before returning to the normal count rate of the observation in question. A full-array timestream with many cosmic events can be seen in Figure \ref{fig:cosmics}. To prevent any possible contamination from cosmic events, 5000 $\mu$s and 10000 $\mu$s are removed from before and after each event, respectively (the asymmetrical nature is due to there being a `tailing off' behavior while the energy spreads through the array). This veto leads to a short time range surrounding each cosmic ray event being removed, ultimately removing $\sim0.3\%$ of the total time each resonator was taking data. 

For the last quality cut we remove any ``hot'' pixels remaining. Qualitatively, a pixel is considered ``hot'' if -- after cosmic rays are removed -- it counts significantly more photons than other pixels. These pixels that register photons more frequently than the average pixel may become "hot" for reasons such as having a very low phase threshold, being under- or overpowered, electrical noise in the readout electronics causing the probe tone for that individual resonator to become noisy, the resonator being adversely affected by local electrical or magnetic fields and shifting slightly in frequency, causing the resonator to go out of calibration, or other more complicated phenomena leading to a specific resonator triggering more frequently than its neighbors. The hot pixels cause the distribution of the total number of photons measured per resonator to be highly positively skewed. To properly characterize the shape of this distribution and catch hot pixels, a metric that is robust to outliers must be chosen. To this end, the median of the number of counts ($\hat{c}$) from each resonator was taken to be the `expected' value of the distribution while the spread is measured using the \texttt{astropy mad\_std} function (\url{https://docs.astropy.org/en/stable/api/astropy.stats.mad_std.html}) to calculate a robust standard deviation ($\sigma_{c, MAD}$) using the Median Absolute Deviation (MAD). As shown in equation \ref{eqn:hotpixel} a pixel is considered hot if the number of counts $c$ it sees is greater than the median number of counts measured by all pixels plus 15 times the MAD standard deviation.

\begin{equation}
\label{eqn:hotpixel}
    c\geq\hat{c}+15\sigma_{c,MAD}
\end{equation}

This cut leaves 590 of the 779 pixels that remained from Section \ref{sub:array_cal}. Ultimately this represents a cut of 24.3$\%$ of the pixels that were energy calibrated at all 5 laser energies (or 26.7$\%$ of the initial 819 pixels).

\begin{figure}
    \centering
    \includegraphics[width=\columnwidth]{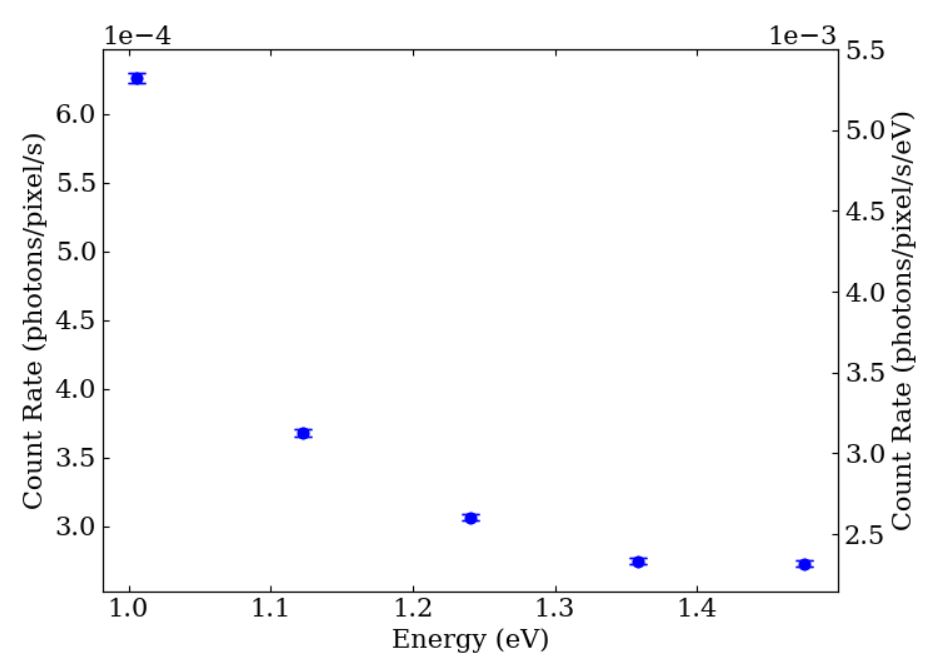}
    \caption{The average count rate values in photon counts per pixel per second in each energy bin with 1$\sigma$ error bars measured from the ensemble of count rates from all remaining pixels.}
    \label{fig:dark_count_rate_preliminary}
\end{figure}

The median resolving power of the remaining 590 MKID pixels across all wavelengths is $\mathcal{R}\sim4.54$. For each pixel the photons which fell between the calibrated energy values and were not removed by the cosmic ray calibration were divided into equal-width energy bins and the count rate of photons per second was then calculated at each energy. This allowed the count rate across pixels to be measured, whose values can be seen in Figure \ref{fig:dark_count_rate_preliminary} and Table \ref{table:countrates_compared_digitalreadout}.

\subsection{Analysis}
\label{sub:digital_analysis}

In Section \ref{sub:digitalreadout_collectionandreduction} the data collection and reduction was discussed in detail. After the final subset of pixels was determined the dark count rate could be calculated. To do this, the count rate in each pixel was measured and the error on each value calculated using Poisson statistics. 

The energy bins were determined by the calibrated bandpass (0.946-1.534 eV, or 1310-808 nm) and the median resolving power $\mathcal{R}\sim5$. This resulted in 5 energy bins of 0.118 eV centered at 1.005, 1.123, 1.240, 1.358, and 1.476 eV.

The measured count rate for the unilluminated MKID array ranges from $(2.73\pm0.02)\times10^{-4}$ photons/pixel/s at 1.476 eV to $(6.26\pm0.04)\times10^{-4}$ photons/pixel/s at 1.005 eV after cosmic ray rejection and other cleaning steps, respectively. This corresponds to an MKID pixel seeing a low energy photon roughly every 1600$\pm$10 seconds and a high energy photon every 3660$\pm$30 seconds. Alternatively, over the full bandpass the count rate is measured at $(1.847\pm0.006)\times10^{-3}$ photons/pixel/s. In flux units, this is $(3.14\pm0.01)\times10^{-3}$ photons/pixel/s/eV in energy space  or $(3.68\pm0.01)\times10^{-6}$ photons/pixel/s/nm in wavelength space.

Although the dark count rate measured by an MKID pixel is not directly analogous to the dark current measured by CCDs and EMCCDs (Electron-Multiplying Charge Coupled Devices) due to the differing origin of the events it is useful to compare the event rates seen by each. Above, we reported that the number of dark count events measured per MKID pixel is $(1.847\pm0.006)\times10^{-3}$ photons/pixel/s. State-of-the-art CCDs have measured dark count rates of $1.66\times10^{-3}$ electrons/pixel/s\cite{CastelloMor2020} while EMCCDs have measured dark current rates as low as $1\times10^{-3}$ electrons/pixel/s\cite{Daigle2012}. Assuming that the CCD and EMCCD gain (the ratio of photons needed to generate 1 electron in the CCD/EMCCD) is 1 -- i.e. 1 electron in the detector corresponds to 1 photon event -- then the dark current generation in these state-of-the-art CCDs and EMCCDs are comparable to the dark count rates measured by an MKID pixel.

\subsubsection{Comparison to raw data}
In Section \ref{sub:digitalreadout_collectionandreduction} the data reduction algorithm was discussed. This includes the removal of photons from cosmic ray events, cleaning of particularly noisy sections of time, and excluding `hot' pixels that live in a state where they count significantly more photons than their physical neighbors. To show the improvement these cuts bring we also calculate the dark count rate with none of the removal steps performed. The results are shown in Table \ref{table:countrates_compared_digitalreadout}. 

Prior to performing any data cleaning, each resonator saw an average of 1493$\pm$2 counts across the calibrated bandpass. Afterwards this number was reduced to  159.5$\pm$0.5, an improvement of by nearly a factor of 10. While we were able to reduce the number of counts by almost a factor of 10 for each energy band, the time cut was not particularly aggressive and only about 0.3$\%$ of the duration from each pixel was removed via the cosmic ray cuts. This shows that the majority of the counts seen come from spurious events that can be calibrated out without significant time needing to be removed from the dataset in question.

\begin{table}
    \centering
    \begin{tabular}{|| c | c  c | c  c ||}
        \hline
        \multicolumn{1}{|| c |}{} & \multicolumn{2}{| c |}{With Reduction Steps} & \multicolumn{2}{| c ||}{No Reduction Steps} \\
        \hline
        Bin Center & Count Rate ($\times10^{-4}$) & Total Counts & Count Rate ($\times10^{-3}$) & Total Counts \\
        (eV) & (photons/pixel/s) & (photons) & (photons/pixel/s) & (photons) \\
        \hline\hline
        1.005 & 6.26$\pm$0.04 & 54.1$\pm$0.3 & 5.0$\pm$0.1 & 436.7$\pm$0.9 \\
        1.123 & 3.68$\pm$0.03 & 31.8$\pm$0.2 & 3.83$\pm$0.09 & 332.2$\pm$0.8 \\
        1.240 & 3.06$\pm$0.02 & 26.5$\pm$0.2 & 3.17$\pm$0.08 & 275.2$\pm$0.7 \\
        1.358 & 2.75$\pm$0.02 & 23.7$\pm$0.2 & 2.73$\pm$0.07 & 237.1$\pm$0.6 \\
        1.476 & 2.73$\pm$0.02 & 23.6$\pm$0.2 & 2.44$\pm$0.07 & 211.6$\pm$0.6 \\
        \hline
    \end{tabular}
    \caption{The average count rate per pixel and total number of counts per pixel for the data with the different calibration and cleaning steps (see also Figure \ref{fig:dark_count_rate_preliminary}) compared to the same quantities without. Without reduction steps, each pixel took data for 86,750 seconds. After cleaning, each pixel was left with 86,491 seconds of data.}
    \label{table:countrates_compared_digitalreadout}
\end{table}

\subsubsection{Potential photon sources}
The steps taken during the data reduction in Section \ref{sub:digitalreadout_collectionandreduction} aimed to mitigate effects from noisy pixels and well-understood noise sources such as cosmic ray events. However there is still the possibility that there are more complex or uncalibrated sources that are not well characterized in this system. This can include things such as electrical noise from cryogenic amplifiers or room temperature readout electronics, secondary photons stemming from cosmic rays exciting electrons and causing fluorescence in the fiber optic cable, and simple blackbody radiation.

Work has been done to characterize the noise characteristics of the MKID Digital Readout\cite{Fruitwala2020} although it is not well understood how electrical noise in the system translates to spurious triggers on non-photon events. However, well-behaved MKID pixels are partially characterized by showing low noise, meaning they will be less susceptible to pixel-specific noise causing a false photon trigger.

In the event that a cosmic ray is absorbed at a point in the fiber optic cable or the rest of the optical path it may deposit its energy in that material, exciting electrons which will then release secondary photons from this particle being absorbed. In this case, individual photons may be generated in the optical path which can then be transmitted to the MKID detectors. This would be a `true' photon detection from an unintended physical source.

The previous two sources of photon detections are both issues that may contaminate sensitive data in a photon-starved environment but are currently challenging if not impossible to mitigate. For the first, reducing electrical noise by preventing ground loops, using low-noise power sources, and working in an isolated environment will reduce false triggers from electronic noise but in practice this is nearly impossible to eradicate. We note however that large scale electrical noise typically affects all resonators simultaneously and can therefore be removed (using a similar time-coincidence veto as a cosmic ray) or causes single resonators to become `hot' or `cold' which may also be handled gracefully in the data reduction pipeline \cite{SteigerAndBailey2022}. For the second, a secondary photon from a cosmic ray may be removed in the data reduction if its energy is sufficiently far outside the calibrated bandpass of the detector, but if its energy is within the bandpass then it will be impossible to remove as it is a single photon event and therefore not subject to the same time-coincidence veto from when a cosmic ray strikes the detector directly.

\begin{figure}
    \centering
    \includegraphics[width=0.8\columnwidth]{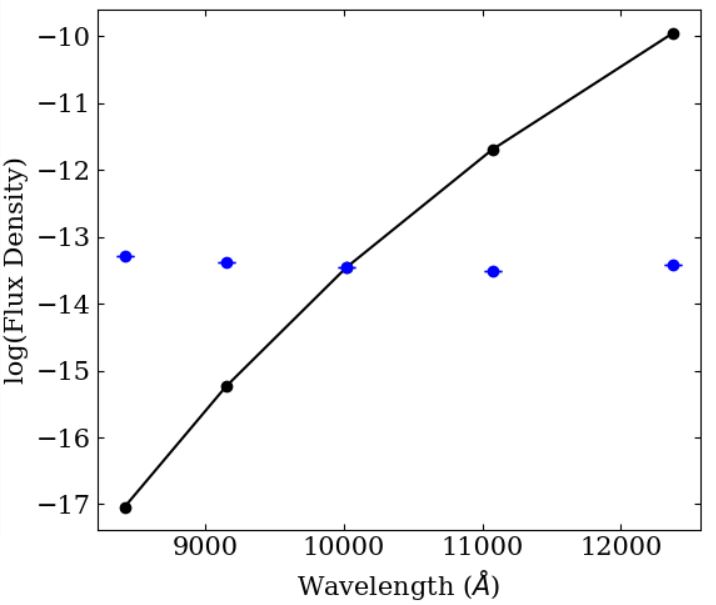}
    \caption{MKID spectrum (blue) compared to a 300 K blackbody spectrum (black) scaled to the central value of the MKID spectrum shown on a logarithmic scale. In this bandpass the blackbody spectrum varies over 7 orders of magnitude while the MKID spectrum remains relatively flat. The central point of the 300 K spectrum is normalized to the central point of the MKID spectrum. Error bars on the MKID spectrum are sufficiently small that they are contained within the points themselves.}
    \label{fig:bb_compare}
\end{figure}

We will explore the possibility that the photons that the MKIDs are seeing while purportedly unilluminated are coming from a blackbody radiating within the dilution refrigerator. Although all precautions have been taken to prevent stray photons from hitting the detector, photons are incredibly difficult to insulate against so we examine the possibility that the photons measured in the dark environment come from a thermal source.

First, we generate blackbody spectra for each of the 4 potential temperature stages which may be generating blackbody photons that could possibly hit the MKIDs. These are the 100 mK stage where the array itself and fiber collimator are mounted and the array is directly exposed to, the walls of the 4 K and 50 K intermediate stages that are used to step down from room to operating temperature and that are nested around the 100 mK stage, and 300 K, which represents the blackbody radiation from the ambient environment or the inner face of the outermost temperature stage of the dilution refrigerator.

Using Planck's Law
\begin{equation}
\label{eqn:plancks_law}
    B_{\lambda}(T)=\frac{2hc^{2}}{\lambda^{5}}\frac{1}{exp\big(\frac{hc}{\lambda k_{B}T}\big)-1}
\end{equation}
we find that the flux density of 100 mK, 4 K, and 50 K blackbody radiation between $\sim$0.9 eV and $\sim$1.6 eV is sufficiently small that a blackbody at any of these temperatures would not be expected to produce photons from 0.9-1.6 eV over the duration of the experiment (86,750 seconds). Therefore, the blackbody spectra representing these temperatures are not included in the analysis.

The 300 K blackbody spectrum can be seen plotted against the spectrum measured by the MKID pixels in Figure \ref{fig:bb_compare}. The spectra are shown in units of log$_{10}$(Flux Density), where the flux density was measured/calculated in ergs/s/cm$^{2}/\mathrm{\mathring{A}}$. The plot is shown in log scale because while the MKID spectrum remains relatively flat across the bandpass, ranging from $(5.22\pm0.04)\times10^{-14}$ ergs/s/cm$^{2}/\mathrm{\mathring{A}}$ to $(3.07\pm0.02)\times10^{-14}$ ergs/s/cm$^{2}/\mathrm{\mathring{A}}$, the 300 K blackbody spectrum varies over 7 orders of magnitude. 

The massive discrepancy in shape of the two spectra show that the photon hits that are still being measured by the MKID pixels are not generated by a 300 K blackbody. With this and the fact that the 100 mK, 4 K, and 50 K stages will not generate any blackbody photons over the calibrated bandpass it is possible to say that the source of the remaining photons measured using the MKID digital readout do not come solely from blackbody sources in the environment.

\section{MKID analog readout: single detector measurements}
\label{sec:analogreadout}

\subsection{Experimental setup}
\label{sub:analogreadout}

The analog readout utilizes the same internal electronics of the fridge, but externally the 6 foot SMA cables attach to a homodyne readout system consisting of two Anritsu MG37022A Signal generators, a Weinschel Attenuator box 8310 Series, an National Instruments-ADC/DAC, and an IQ mixer box. The function of these devices is the same as in the digital readout case. The schematic for the analog readout system, and the parametric amplifier, is shown in Figure \ref{fig:dilution_refrigerator}. Unlike the digital readout, the analog readout supplies individual frequencies from an Anritsu Synthesizer to probe single resonators on the array. The analog readout has less noise associated with it compared to the digital readout which has to make compromises so that it is able to issue many probe tones while also dealing with limited dynamic range in the ADC attenuators and precision in its firmware computations.

The primary reason for taking a set of data in the dark with the analog readout is to analyze the nature of the photon pulses. Despite the fact that MKIDs are not susceptible to read noise and dark current due to their unique detection mechanism compared to conventional detectors such as CCDs\cite{day2003, Mazin2012, Fruitwala2020}, empirical evidence has shown that they do indeed still measure photon-like events when they are not illuminated that may be triggered by noise sources such as those noted in Section \ref{sub:mkid_array}. Since the MKID Analog Readout saves photon pulse timestream data (as in Figures \ref{fig:photon_timestream} and \ref{fig:bad_photon_triggers}) it is possible to explore the nature of these events to determine if these `dark counts' look the same as `true' photon events, or if they are demonstrably different. By measuring several thousand `dark' photons this way we aim to assess them and determine if we can assign them any explainable origin.

\subsection{Data collection and reduction}
A second set of data was collected in addition that from the MKID Digital Readout while the MKID was unilluminated where an MKID pixel was read out using an analog readout system designed for low noise, single-pixel characterization. 

We chose a single MKID resonator that was also read out using the digital readout that had above-average resolving powers at all calibration energies. In this configuration there is only one MKID pixel read out, so we are no longer able to leverage the bulk properties of the MKID array for cosmic ray rejection. However, the analog readout saves the phase timestream of the resonator surrounding each photon event which allows inspection of each pulse (see Figures \ref{fig:photon_timestream} and \ref{fig:bad_photon_triggers}) to determine whether it is characteristic of a `real' photon or whether noise caused the trigger.

In contrast to the digital readout which continuously takes data until the user decides to stop, the analog readout accepts the number of desired photon counts to measure before stopping. In this case the quiescent count rate of photon counts in the dark was first measured and found to be $\sim$0.03 Hz, which likely consists of predominantly cosmic ray events. With this in mind we chose to register 8000 photon counts and expected this should take roughly 3 days. The primary goal of this investigation is to see if the photons which are being triggered on look `real' or if they look like noise, although it is also possible to ascertain a dark count rate.


The analog readout system saves its data in a slightly different structure than the digital readout of Section \ref{sec:digitalreadout}. The digital readout saves time-tagged lists of photons along with the pixel location and height of the photon event but due to computational constraints do not save further information about the photons. In contrast, the analog readout operates in a way so that the recent phase data from each MKID pixel being read out is kept in memory so that when a photon is measured, the readout system may save a timestream of that phase data from the pixel from the time surrounding the photon event. An example of a phase timestream saved by the analog readout is shown in Figure \ref{fig:photon_timestream}. The duration and sample rate of this phase timestream are both parameters that can be tuned by the user. In this experiment the resonator's phase was sampled at 0.8 MHz and each phase timestream saved the 5000 $\mu$s surrounding the photon event (2500 $\mu$s before and after).

Analogously to the MKID Science Data pipeline, we first reject any photons which are outside of the calibrated bandpass (i.e. the peak of the phase is too high or too low) as well as any timestreams that contain more than 1 photon hit. The second criterion is the closest proxy we have to a time-coincidence veto in lieu of using bulk statistics from many pixels. This leaves 1118 photons, $\sim14\%$ of the total observed counts. In comparison, the cuts from the digital readout left us with $\sim10.7\%$ of the total observed counts (94262 remained of the initial 880855 from the analyzed pixels). The percentage for the analog readout is much due to both the lower noise from the readout system resulting in fewer counts below the calibrated region (i.e. triggers on noise) as well as the inability to make simultaneity cuts on cosmic ray events in the analog case.

\subsection{Analysis}
By binning photons using the same energy bins as in the MKID Digital readout we are able to compare the improvement that is gained when reading out a single pixel using significantly less noisy readout electronics. The count rates within the calibrated bandpass are shown in Table \ref{table:digital_to_analog_compare}. As in the digital readout analysis the errors on photon counts and ultimate count rates are calculated using Poisson statistics.

\begin{table}
    \centering
    \begin{tabular}{|| c | c  c | c  c ||}
        \hline
        \multicolumn{1}{|| c |}{} & \multicolumn{2}{| c |}{Analog Readout} & \multicolumn{2}{| c ||}{Digital Readout (No Reduction)} \\
        \hline
        Bin Center & Count Rate ($\times10^{-3}$) & Total Counts & Count Rate ($\times10^{-3}$) & Total Counts \\
        (eV) & (photons/pixel/s) & (photons) & (photons/pixel/s) & (photons) \\
        \hline\hline
        1.005 & 1.8$\pm$0.1 & 189$\pm$14 & 5.0$\pm$0.1 & 436.7$\pm$0.9 \\
        1.123 & 1.4$\pm$0.1 & 148$\pm$12 & 3.83$\pm$0.09 & 332.2$\pm$0.8 \\
        1.240 & 1.4$\pm$0.1 & 145$\pm$12 & 3.17$\pm$0.08 & 275.2$\pm$0.7 \\
        1.358 & 1.4$\pm$0.1 & 141$\pm$12 & 2.73$\pm$0.07 & 237.1$\pm$0.6 \\
        1.476 & 1.2$\pm$0.1 & 121$\pm$11 & 2.44$\pm$0.07 & 211.6$\pm$0.6 \\
        \hline
    \end{tabular}
    \caption{Comparison of the total counts in the calibrated bandpass between 0.946-1.534 eV (1310-808 nm) using the MKID analog readout scheme to the MKID digital readout system. }
    \label{table:digital_to_analog_compare}
\end{table}

\subsubsection{Photon rise and fall times}
As previously discussed, the analog readout system saves photon timestream data (such as those shown in Figures \ref{fig:photon_timestream} and \ref{fig:bad_photon_triggers}). This allows us to examine the characteristic rise and fall times of the MKID pixel's phase when a photon event is triggered. For a baseline measurement an 808 nm (0.946 eV) laser and a 1310 nm (1.534 eV) laser are each shined on the pixel until it has registered 20,000 photon events.

A photon event is characterized by a fast exponential rise time in the measured phase as a photon strikes the detector, depositing its energy and breaking Cooper pairs into quasiparticles followed by a slower exponential tail as the quasiparticles recombine. To fit the rise and the fall times for a given photon event, the timestream is first split into a rising portion from the start of the timestream to the peak and a falling portion from the peak to the end. The rise and fall times are then each calculated from their respective sections of the timestream by fitting an exponential of the form

\begin{equation}
    \phi = A e^{-(t/b)} + c
\end{equation}
where $\phi$ is the measured phase, $t$ is the corresponding time within the photon timestream, $b$ is the time constant (which we call the rise or the fall time depending on which part of the event we are fitting), and $A$ and $c$ are constants to account for any offset or scaling differences between pulses.

\begin{figure}
    \centering
    \includegraphics[width=\columnwidth]{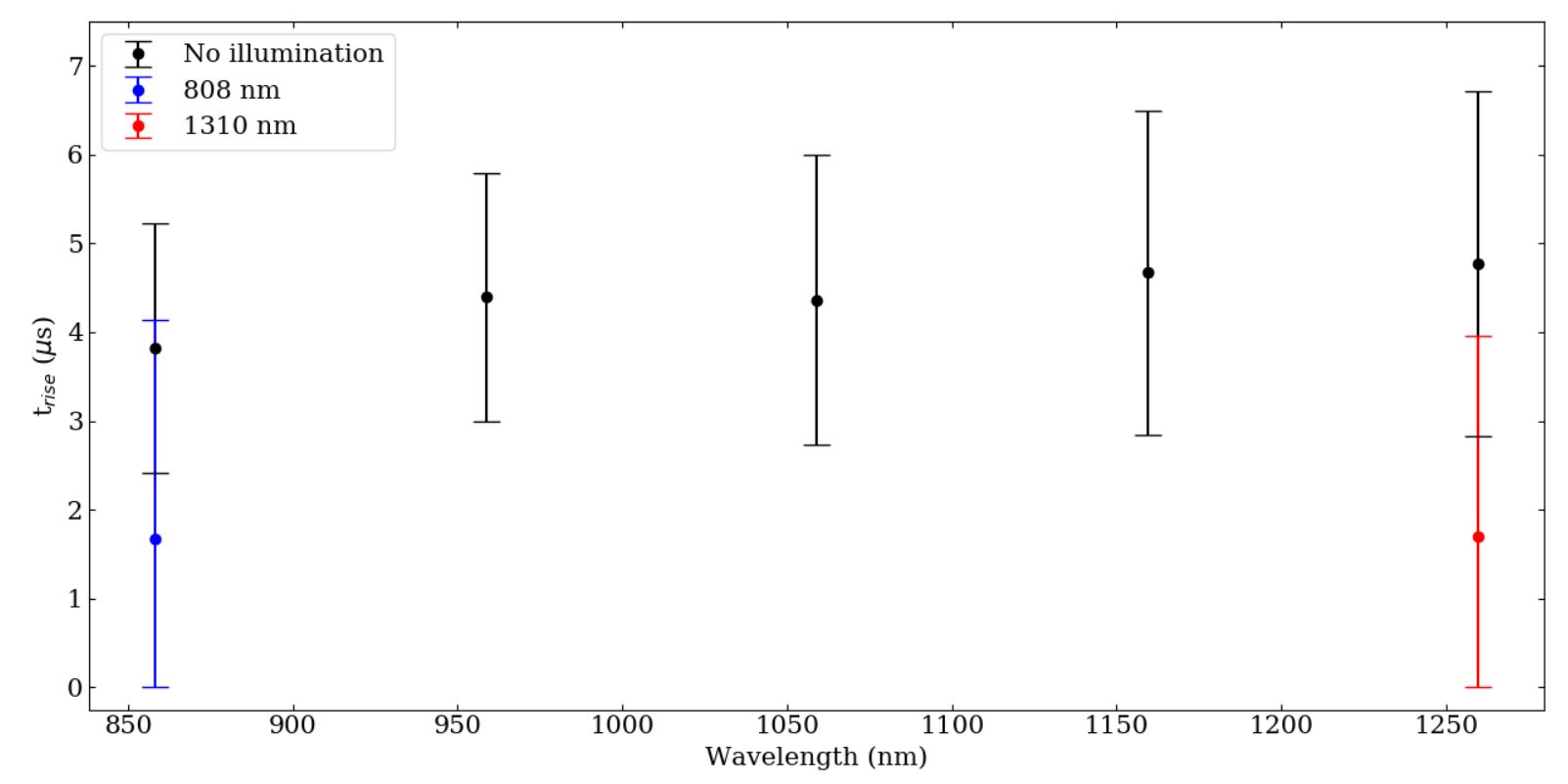} \\
    \includegraphics[width=\columnwidth]{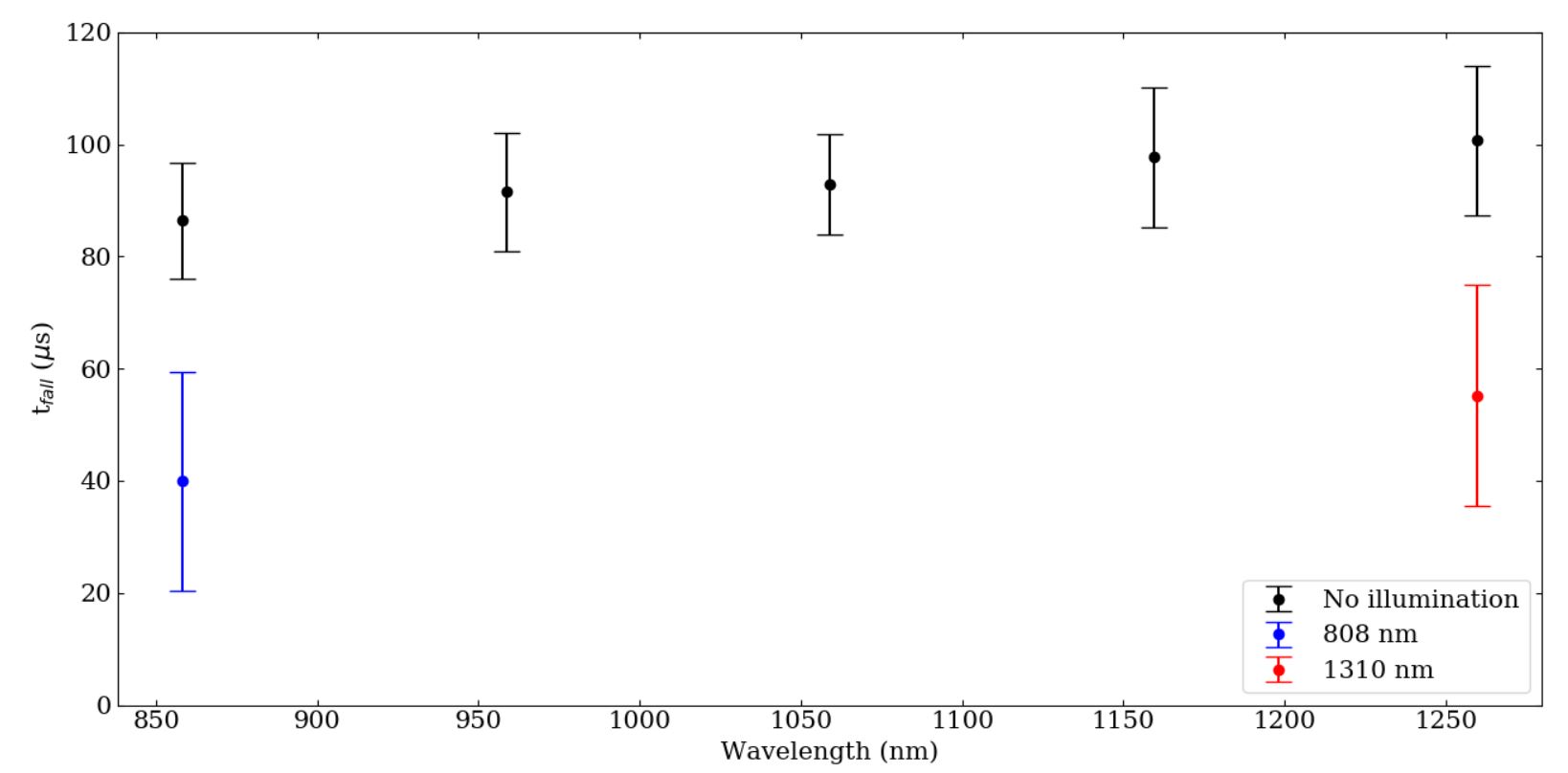}
    \caption{Comparison of the rise times (top) and fall times (bottom) as a function of photon wavelength between dark counts - shown in black - and photons from 808 and 1310 nm lasers - shown in blue and red, respectively. The error bars for each point are the 1$\sigma$-errors for each measurement. In the top panel it can be clearly seen that the rise times of dark count photons and laser photons overlap significantly. In the bottom panel, the fall times for the dark count events do not overlap the photon events from the lasers.}
    \label{fig:photon_timescales}
\end{figure}

\begin{figure}
    \centering
    \includegraphics[width=\columnwidth]{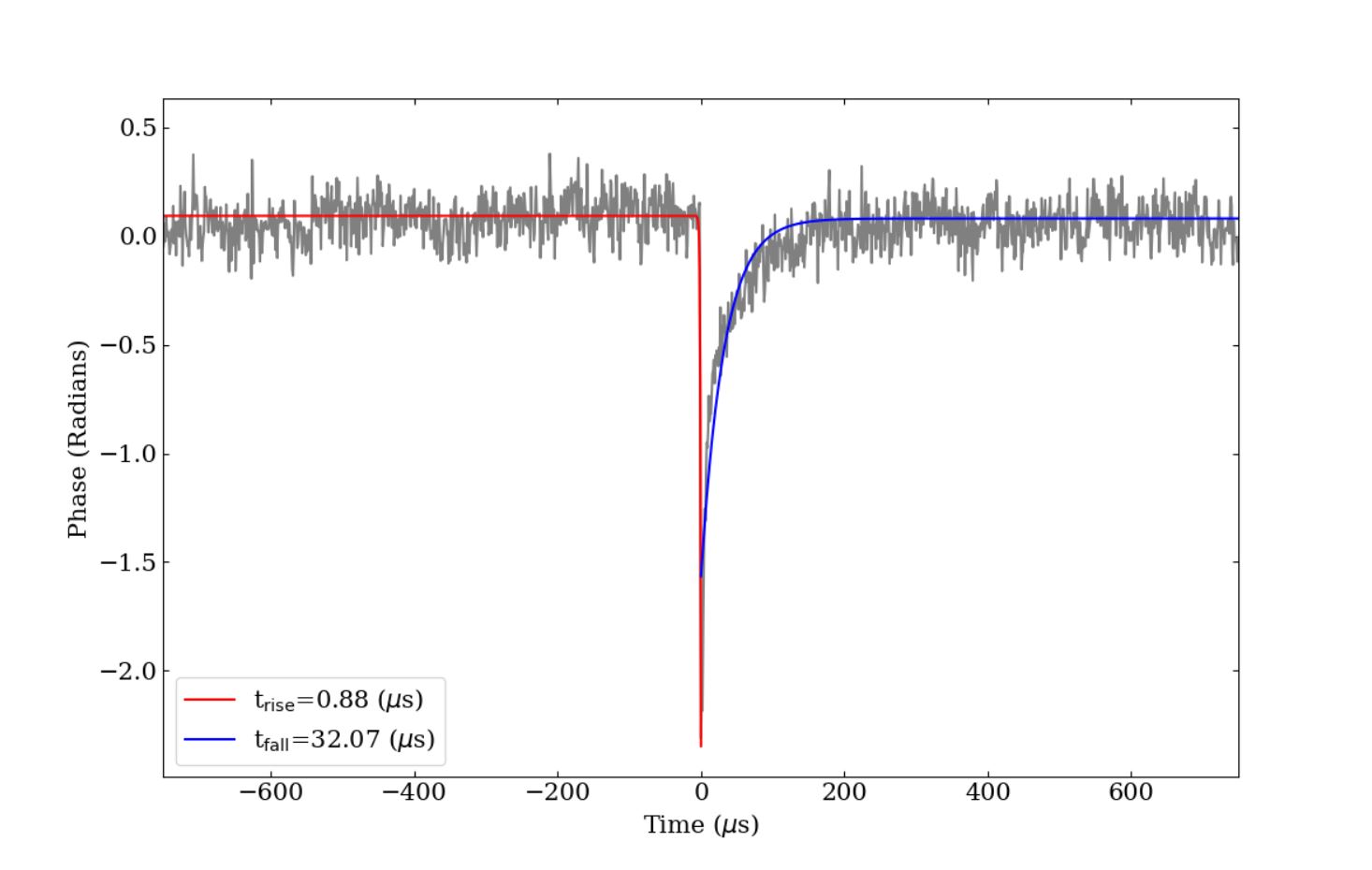}
    \includegraphics[width=\columnwidth]{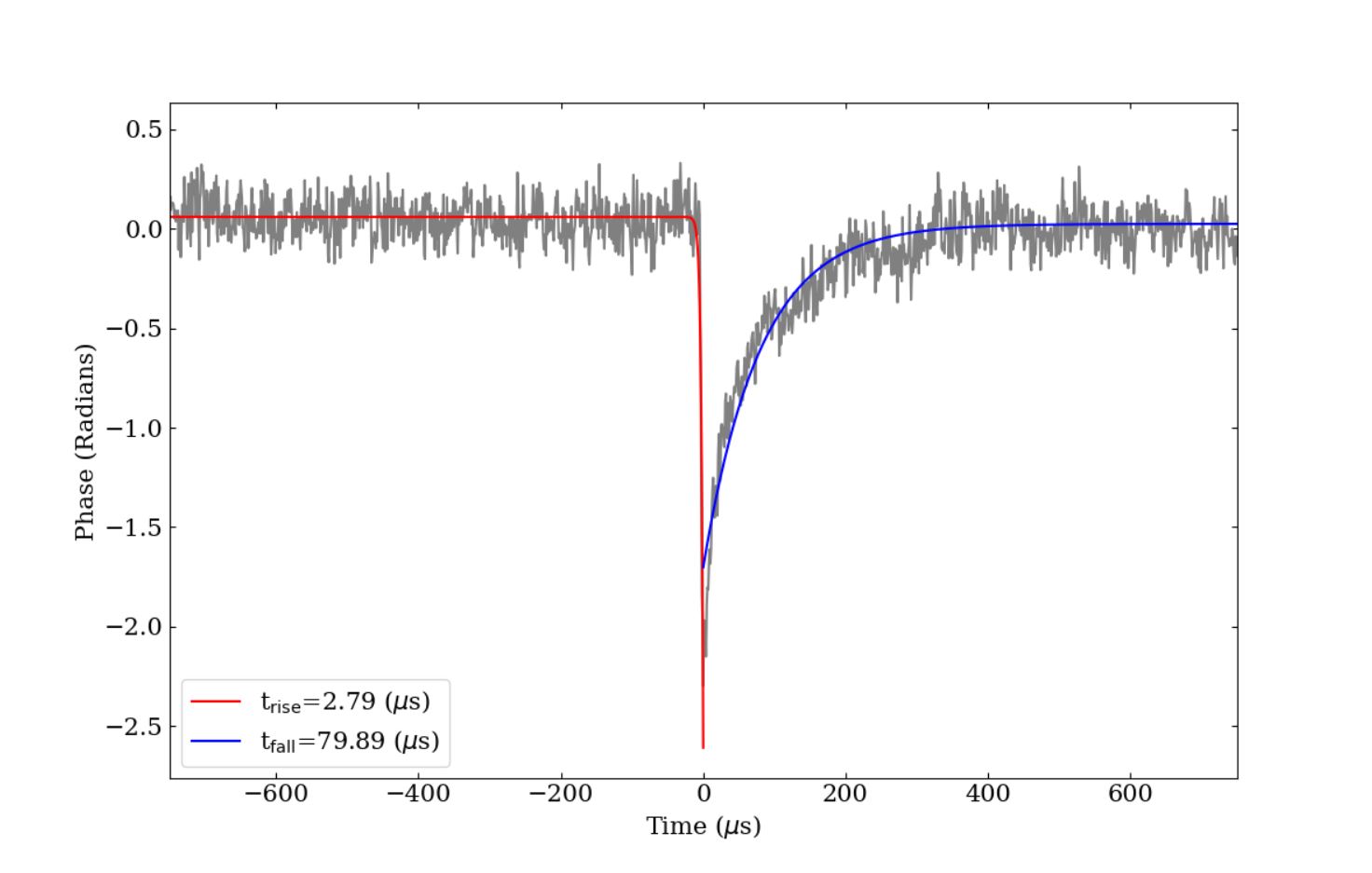}
    \caption{Comparing the calculated rise and fall times of photons of the same energy measured by the MKID resonator. (Top) An 808 nm photon measured with an 808 nm laser shining on the pixel. (Bottom) An 808 nm photon measured when there is no light incident on the pixel. This corroborates the significant difference in fall times shown in Figure \ref{fig:photon_timescales}.}
    \label{fig:rise_vs_fall_times}
\end{figure}

We fit the rise and fall times for all of the photons from both the 808 and 1310 nm lasers as well as the dark-count photons measured by the pixel while it was unilluminated. Figure \ref{fig:photon_timescales} shows how the rise and fall times of the dark counts compare to those of known photons from the two lasers as a function of wavelength. 

It can be clearly seen that the 1$\sigma$ (68$\%$) confidence intervals of the rise times of the dark count photons overlaps with the 1$\sigma$ CI of the rise times of the 808 and 1310 nm photons. This is relatively unsurprising as phase spikes so sharply when energy is deposited into the pixel that it is effectively instantaneous even with the microsecond timing resolution.

On the other hand, when the fall times of the dark counts are compared to those of the photons from the lasers one can see a stark distinction between the two distributions. The fall times from the laser photons are significantly faster than those of the dark counts for photons of similar energy - and therefore similar phase response. In theory the phase decay back to its quiescent value is governed by the quasiparticle recombination time, an effect solely due to properties of the superconducting material. Since the phase response is proportional to the number of quasiparticles generated in the material, photons that cause similar phase responses should have similar fall times since a roughly equivalent number of quasiparticles need to combine back into Cooper pairs. We can see in the bottom panel of Figure \ref{fig:photon_timescales} that this is not the case in this experiment. Figure \ref{fig:rise_vs_fall_times} shows an example of each of these photons. In the top panel, a photon that was measured when a laser was incident on the pixel can be seen with a measured fall time of 32.07 $\mu$s. Below that a second photon of the same energy can be seen, this time from when there was no light incident on the array. The fall time in this case is measured to be 79.89 $\mu$s, consistent with the difference in the distribution of the fall times from each population. 

Although the dark count photons that remain after all previous cuts \textit{qualitatively} appear similar to the photons measured when lasers were being shined on the MKIDs, there is a quantitative difference in the two populations. The photons that are measured when a laser is being shined on the MKID show signficantly shorter fall times than dark count photons measured by the same pixel. The explanation for this behavior stems from the sources of the different families of photons and how they deposit energy into and are subsequently measured by the pixel. The rise and fall times of a photon event correlates with how quickly energy is deposited into the resonator and how long it takes for that energy to dissipate and allow the resonator return to its unexcited state. When energy is deposited into the resonator over a short time it will rapidly break many Cooper pairs which then begin to recombine right away, allowing the resonator to return to its unexcited state in a short time. If energy is deposited into the resonator over a longer time scale it will still break many Cooper pairs quickly but as the energy remains in the system before dissipating it will prevent those Cooper pairs from combining as fast which leads to a longer fall time back to quiescence. Real photons from a laser or other light source fall into the first category; the photon is absorbed and all of its energy is immediately deposited into the MKID, leading to a fast rise and fall time. Dark counts from sources such as cosmic rays are in the second family. Photon events triggered by a cosmic ray occur when the energy from the incident energetic particle is down-converted into a cloud of phonons that then spread the absorbed energy through the array substrate. As these phonons move past MKID pixels they deposit some of that energy into the pixel over the time it takes for them to move through the resonator. This means that dark count photons from cosmic ray events are non-instantaneous process and will consequently have longer fall times.

A separate potential explanation for the discrepancy in fall times between the laser photons and dark count photons is that while illuminating the array with a laser the photon flux is significantly higher than when the laser is off. This higher photon flux may lead to slight heating of the thin film of the MKID array. In turn, this would lead to shorter quasiparticle recombination times and ultimately shorter fall times for each photon event. 

There is currently insufficient evidence to conclusively say that these are two completely different populations of photons due to the current gap in understanding of the noise sources in the MKIDs and inability to simultaneously read out an array using both the digital and analog readouts to correlate cosmic ray events (digital) to single photon traces (analog) so we cannot calibrate these photons out based solely on the difference in their fall times than is expected. If the assumption is made that these counts do come from cosmic rays and a calibration cut is made, the number of counts would decrease significantly from 121 to 10 photons in the high energy bin and from 189 to 27 photons in the lowest energy bin, with each bin seeing a reduction by about a factor of 8. Across the full bandpass the count rate would fall from $(7.1\pm0.3)\times10^{-3}$ photons/pixel/s to $(9.3\pm0.9)\times10^{-4}$ photons/pixel/s. In wavelength space this corresponds to $(1.42\pm0.05)\times10^{-5}$ photons/pixel/s/nm and $(1.9\pm0.2)\times10^{-6}$ photons/pixel/s/nm for the `uncalibrated' and `calibrated' rates, respectively. The final value with long-fall time events removed is just slightly below the $(2.77\pm0.02)\times10^{-6}$ photons/pixel/s/nm from the digital readout in Section \ref{sub:digital_analysis}. 

\section{Discussion and conclusions}
In this paper we show that across the calibrated MKID bandpass from 0.946 to 1.534 eV (1310-808 nm), the count rate seen by the detectors in a large format array is $(3.14\pm0.01)\times10^{-3}$ photons/pixel/s/eV or $(3.68\pm0.01)\times10^{-6}$ photons/pixel/s/nm. It as also demonstrated that by using a relatively light calibration cut for cosmic ray events we are able to reduce the number of spurious photon events by nearly a factor 10 of while removing less than 1$\%$ of the duration of the data collection.

Using the MKID analog readout system and recording the shape of photon pulses in a single pixel we first show that the count rate across the calibrated bandpass is $(1.42\pm0.05)\times10^{-5}$ photons/pixel/s/nm without any further data cleaning steps, demonstrating that a quieter system does lead to lower count rates in an unilluminated MKID device.

While investigating the shape of the photon pulses using the analog readout it was found that the exponential tail of the dark counts corresponds to a significantly longer fall time than from photons generated from a laser at the same wavelengths. The long fall times are indicative of energy taking a long time to dissipate from the resonators which points to events causing these triggers happening in the substrate rather than the pixels themselves. An example of a known event that takes place in the substrate and causes contaminating photon events is a cosmic ray hitting the MKID array. Making MKID pixels atop membranes is an ongoing field of research that offers a straightforward way to to minimize the potential for substrate absorptions to cause contaminating photon events. Because of the added complexity of making MKIDs on membranes in addition to the existing difficulty in fabrication this is not feasible for the large-format MKID arrays currently in use. However, in future experiments requiring much fewer (1 to $\sim$100) MKIDs, making them on membranes may be a reasonable path forward that will help mitigate contamination from substrate absorptions. Additionally, the current generation of MKID readout hardware does not allow for side-by-side simultaneous readout of many pixels while still capturing the photon phase timestreams and so there is no way to determine if these long fall time photons are coexistent with cosmic ray events at present. The next 3rd Generation MKID Digital Readout is currently under development \cite{Smith2022} and promises to allow both capabilities at the same time. Further investigation of the source of these dark photons and if they are in fact generated cosmic rays will be explored in future work with the 3rd Generation MKID Readout. We note here that if future work does find that these long fall time photons are from contaminating sources it offers a straightforward way to calibrate them out of MKID datasets.

For a future dark matter detector experiment that would use a 100 pixel MKID array with 10 nm energy bins, the maximum dark count rate in the detector would be $\sim (3.68\pm0.01)\times10^{-3}$ photons/s if the current style arrays and generation of MKID digital readout were used. With this said, any future MKID dark matter direct detection instrument will have several key upgrades to mitigate noise in the system. First, a new generation of MKID readout is currently under development which promises to be a significantly less noisy system than the one used at present. The continued development of Traveling Wave Parametric Amplifiers (TWPA \cite{Eom2012, Zobrist2019}) will also significantly reduce system noise compared to the more commonly used HEMT amplifiers. Finally, this instrument itself will use an array that has anti-reflection (AR) coating on the MKID devices and will not have optics that allow visible light to enter the cryostat. Both of these upgrades will prevent more stray photons from entering the fridge and causing spurious, unattributable counts on the detector.

\section{Acknowledgements}
NS gratefully acknowledges support from the Heising-Simons Foundation under grants $\#$2020-1820 and $\#$2021-3058, as well as support by NASA under grant $\#$80NSSC19K0329 and from the NSF MRI Award $\#$1625441. WHC is grateful for the support from NSTGRO Grant $\#$80NSSC21K1290.

\section{Disclosures}
The authors declare no conflicts of interest. \newline

\noindent\textbf{Data availability.} Data underlying the results presented in this paper are not publicly available at this time but may be obtained from the authors upon reasonable request.

\bibliography{main}

\begin{thebibliography}{10}
\newcommand{\enquote}[1]{``#1''}

\bibitem{day2003}
P.~Day, H.~Leduc, B.~Mazin, A.~Vayonakis, and J.~Zmuidzinas, \enquote{A
  broadband superconducting detector suitable for use in large arrays,}
  {\protect\JournalTitle{Nature}} \textbf{425}, 817--21 (2003).

\bibitem{Szypryt2017}
P.~Szypryt, S.~R. Meeker, G.~Coiffard, N.~Fruitwala, B.~Bumble, G.~Ulbricht,
  A.~B. Walter, M.~Daal, C.~Bockstiegel, G.~Collura, N.~Zobrist, I.~Lipartito,
  and B.~A. Mazin, \enquote{Large-format platinum silicide microwave kinetic
  inductance detectors for optical to near-ir astronomy,}
  {\protect\JournalTitle{Optics Express}} \textbf{25}, 25894 (2017).

\bibitem{Zobrist2019}
N.~Zobrist, B.~H. Eom, P.~Day, B.~A. Mazin, S.~R. Meeker, B.~Bumble, H.~G.
  LeDuc, G.~Coiffard, P.~Szypryt, N.~Fruitwala, I.~Lipartito, and
  C.~Bockstiegel, \enquote{Wide-band parametric amplifier readout and
  resolution of optical microwave kinetic inductance detectors,}
  {\protect\JournalTitle{Applied Physics Letters}} \textbf{115}, 042601 (2019).

\bibitem{Fruitwala2020}
N.~Fruitwala, P.~Strader, G.~Cancelo, T.~Zmuda, K.~Treptow, N.~Wilcer,
  C.~Stoughton, A.~B. Walter, N.~Zobrist, G.~Collura, I.~Lipartito, J.~I.
  Bailey, and B.~A. Mazin, \enquote{Second generation readout for large format
  photon counting microwave kinetic inductance detectors,}
  {\protect\JournalTitle{Review of Scientific Instruments}} \textbf{91}, 124705
  (2020).

\bibitem{Zobrist2021}
N.~Zobrist, N.~Klimovich, B.~H. Eom, G.~Coiffard, M.~Daal, N.~Swimmer,
  S.~Steiger, B.~Bumble, H.~G. LeDuc, P.~Day, and B.~A. Mazin,
  \enquote{Improving the dynamic range of single photon counting kinetic
  inductance detectors,} {\protect\JournalTitle{Journal of Astronomical
  Telescopes, Instruments, and Systems}} \textbf{7} (2021).

\bibitem{Zobrist2022}
N.~Zobrist, W.~H. Clay, G.~Coiffard, M.~Daal, N.~Swimmer, P.~Day, and B.~A.
  Mazin, \enquote{Membraneless phonon trapping and resolution enhancement in
  optical microwave kinetic inductance detectors,}
  {\protect\JournalTitle{Physical Review Letters}} \textbf{129} (2022).

\bibitem{Steiger2021}
S.~Steiger, T.~Currie, T.~D. Brandt, and et~al., \enquote{Scexao/mec and charis
  discovery of a low-mass, 6 au separation companion to hip 109427 using
  stochastic speckle discrimination and high-contrast spectroscopy,}
  {\protect\JournalTitle{The Astronomical Journal}} \textbf{162}, 44 (2021).

\bibitem{Swimmer_2022}
N.~Swimmer, T.~Currie, S.~Steiger, G.~M. Brandt, T.~D. Brandt, O.~Guyon,
  M.~Kuzuhara, J.~Chilcote, T.~Tobin, T.~D. Groff, J.~Lozi, J.~I.~I. Bailey,
  A.~B. Walter, N.~Fruitwala, N.~Zobrist, J.~P. Smith, G.~Coiffard, R.~Dodkins,
  K.~K. Davis, M.~Daal, B.~Bumble, S.~Vievard, N.~Skaf, V.~Deo, N.~Jovanovic,
  F.~Martinache, M.~Tamura, N.~J. Kasdin, and B.~A. Mazin, \enquote{{SCExAO}
  and keck direct imaging discovery of a low-mass companion around the
  accelerating f5 star {HIP} 5319,} {\protect\JournalTitle{The Astronomical
  Journal}} \textbf{164}, 152 (2022).

\bibitem{Steiger_2022}
S.~Steiger, T.~D. Brandt, O.~Guyon, N.~Swimmer, A.~B. Walter, C.~Bockstiegel,
  J.~Lozi, V.~Deo, S.~Vievard, N.~Skaf, K.~Ahn, N.~Jovanovic, F.~Martinache,
  and B.~A. Mazin, \enquote{Probing photon statistics in adaptive optics images
  with {SCExAO}/{MEC},} {\protect\JournalTitle{The Astronomical Journal}}
  \textbf{164}, 186 (2022).

\bibitem{Mazin2012}
B.~A. Mazin, B.~Bumble, S.~R. Meeker, K.~O’Brien, S.~McHugh, and E.~Langman,
  \enquote{A superconducting focal plane array for ultraviolet, optical, and
  near-infrared astrophysics,} {\protect\JournalTitle{Optics Express}}
  \textbf{20}, 1503 (2012).

\bibitem{SteigerAndBailey2022}
S.~Steiger, J.~I. Bailey, N.~Zobrist, N.~Swimmer, R.~Dodkins, K.~K. Davis, and
  B.~A. Mazin, \enquote{The mkid science data pipeline,}  (2022).

\bibitem{Walter2020}
A.~B. Walter, N.~Fruitwala, S.~Steiger, and et~al., \enquote{The mkid exoplanet
  camera for subaru scexao,} {\protect\JournalTitle{Publications of the
  Astronomical Society of the Pacific}} \textbf{132}, 125005 (2020).

\bibitem{Wethimer2011}
D.~Werthimer, \enquote{The casper collaboration for high-performance open
  source digital radio astronomy instrumentation,} in \emph{2011 XXXth URSI
  General Assembly and Scientific Symposium,}  (2011), pp. 1--4.

\bibitem{Hickish2016}
J.~Hickish, Z.~Abdurashidova, Z.~Ali, K.~D. Buch, S.~C. Chaudhari, H.~Chen,
  M.~Dexter, R.~S. Domagalski, J.~Ford, G.~Foster, D.~George, J.~Greenberg,
  L.~Greenhill, A.~Isaacson, H.~Jiang, G.~Jones, F.~Kapp, H.~Kriel, R.~Lacasse,
  A.~Lutomirski, D.~MacMahon, J.~Manley, A.~Martens, R.~McCullough, M.~V.
  Muley, W.~New, A.~Parsons, D.~C. Price, R.~A. Primiani, J.~Ray, A.~Siemion,
  V.~Van~Tonder, L.~Vertatschitsch, M.~Wagner, J.~Weintroub, and D.~Werthimer,
  \enquote{A decade of developing radio-astronomy instrumentation using casper
  open-source technology,}  (2016).

\bibitem{Eom2012}
B.~H. Eom, P.~K. Day, H.~G. Leduc, and J.~Zmuidzinas, \enquote{A wideband,
  low-noise superconducting amplifier with high dynamic range,}  (2012).

\bibitem{CastelloMor2020}
N.~Castell{\'{o}}-Mor, \enquote{{DAMIC}-m experiment: Thick, silicon {CCDs} to
  search for light dark matter,} {\protect\JournalTitle{Nuclear Instruments and
  Methods in Physics Research Section A: Accelerators, Spectrometers, Detectors
  and Associated Equipment}} \textbf{958}, 162933 (2020).

\bibitem{Daigle2012}
O.~{Daigle}, O.~{Djazovski}, D.~{Laurin}, R.~{Doyon}, and {\'E}.~{Artigau},
  \enquote{{Characterization results of EMCCDs for extreme low-light imaging},}
  in \emph{High Energy, Optical, and Infrared Detectors for Astronomy V,}  vol.
  8453 of \emph{Society of Photo-Optical Instrumentation Engineers (SPIE)
  Conference Series} A.~D. {Holland} and J.~W. {Beletic}, eds. (2012), p.
  845303.

\bibitem{Smith2022}
J.~P. Smith, J.~I. Bailey, and B.~A. Mazin, \enquote{Highly-multiplexed
  superconducting detector readout: Approachable high-speed fpga design,} in
  \emph{2022 IEEE 30th Annual International Symposium on Field-Programmable
  Custom Computing Machines (FCCM),}  (2022), pp. 1--2.

\end{thebibliography}

\end{document}